# Spintronics with graphene

Pierre Seneor, Bruno Dlubak, Marie-Blandine Martin, Abdelmadjid Anane, Henri Jaffres, and Albert Fert

*Unité Mixte de Physique CNRS/Thales, 91767 Palaiseau, France and Université de Paris-Sud 11, 91405 Orsay, France*

**Because of its fascinating electronic properties, graphene is expected to produce breakthroughs in many areas of nanoelectronics. For spintronics, its key advantage is the expected long spin lifetime, combined with its large electron velocity. In this article, we review recent theoretical and experimental results showing that graphene could be the long-awaited platform for spintronics. A critical parameter for both characterization and devices is the resistance of the contact between the electrodes and the graphene, which must be large enough to prevent quenching of the induced spin polarization but small enough to allow for the detection of this polarization. Spin diffusion lengths in the 100-µm range, much longer than those in conventional metals and semiconductors, have been observed. This could be a unique advantage for several concepts of spintronic devices, particularly for the implementation of complex architectures or logic circuits in which information is coded by pure spin currents.**

## Introduction

With hundreds of millions of computer hard drives sold every year, magnetism is currently, by far, the main repository of information storage. This dominance will only increase with the expected proliferation of data centers for "cloud" access over the Internet. It is the electron "spin," the elementary nanomagnet, that carries this information. Beyond storage, spin is foreseen as the foundation for a new paradigm for information processing toward low-power-consumption nonvolatile "green" electronics. This is the aim of spintronics.

However, despite intense research, a simple device such as the spin transistor proposed in 1990[1] has remained elusive. Whereas it was soon realized that fundamental constraints on the physics of spin transport would make this concept very difficult to achieve with conventional semiconductors such as GaAs or silicon[2] (indeed, electrical injection of a spin current directly into silicon was demonstrated only recently[3]), a suitable material was still sought.

Recently, because of its fascinating electronic properties, graphene has become the focus of expectations for producing breakthroughs in many areas of nanoelectronics.[4–8] For spintronics, graphene's obvious attraction is mainly the long spin lifetime expected from the small spin–orbit coupling of carbon atoms and the absence of nuclear spins for the main isotope. The combination of this expected long spin lifetime with a high electron velocity, related to the linear dispersion relation of electrons in graphene, underlies the potential of graphene for spintronics. The ability to transport spin information efficiently over practical distances could further enable complex spintronic devices, such as the reconfigurable logic gate integrating both memory and logic proposed by Dery et al.,[9] and eventually open the way to spin information processing.

The spin diffusion distances observed in graphene are very long, in the 100-µm range, much longer than those in conventional metals and semiconductors. This is a unique advantage for several concepts of spintronic devices, particularly for complex architectures in which information is coded by pure spin currents and processed by series of logic gates acting on their spin polarization.

Indeed, although a suitable platform for such devices remains to be identified, initial steps in this direction have already been taken, such as non-charge-based "beyond-CMOS" memory and logic devices highlighted in the Emerging Research Devices chapter of the International Technology Roadmap for Semiconductors (ITRS).[10] Among several other spintronic devices, so-called "all-spin-logic" circuits based on the transport and processing of information coded by spin currents have been proposed.[11]

Research on a suitable platform for spin transport started decades ago and focused first on conventional inorganic semiconductors and metals. However, these materials have shown limited spin signals and/or spin diffusion lengths that are typically in the range of only a few tenths of a micrometer at room temperature. Graphene, in contrast, appears to be potentially well adapted for the transport of spin information over relatively long distances in the 100-µm range with limited spin losses. Some organic molecules or carbon nanotubes could also be expected to provide this performance, but graphene is more convenient for practical devices.

Although the injection and detection processes still need to be improved and the relaxation mechanisms need to be understood, it has been demonstrated that spin-polarized currents in graphene can give rise to large electrical signals. In this article, we describe these experiments and the theoretical framework that enables their interpretation and optimization. In particular, matching the resistance of the tunneling contacts to spin relaxation in graphene is key to achieving efficient spin transport.

*Early results on exfoliated graphene*

The first experiment on spin transport in graphene was reported in 2006, for a graphene flake connected to NiFe electrodes.[12] The device was a lateral spin valve (LSV), in which a spin polarized current injected from one ferromagnetic electrode travels through the graphene before being detected by a second electrode. In analogy to optical systems, the ferromagnetic electrodes act as a polarizer/analyzer set. The current flowing through the device depends on the persistence of the spin polarization and whether the electrodes are magnetically aligned in a parallel or antiparallel configuration. This measurement was rapidly followed by several other spin-transport measurements on single-layer graphene (SLG) and multilayer graphene (MLG).[13–16] Mainly, two configurations are reported for the measurements: one "local" and the other "nonlocal". The local configuration is a simple two-terminal device acting as an LSV. In the nonlocal configuration, four terminals are used, in a geometry slightly different from that used in conventional fourpoint measurements. Specifically, as seen in Figure 1, the current path is separated from the voltage measurement zone. The nonlocal configuration was originally developed to extract low signals[18] in semiconductors and metals, where the non-spin-aligned current would overwhelm the signal in local LSV measurements.

As an example of this nonlocal technique, Figure 1a shows the SLG device studied by Popinciuc et al.[17] Injecting a spin-polarized current at electrode 2 (Figure 1b) creates an out-of-equilibrium spin population in the graphene layer.

The difference between the electrochemical potentials of the spin-up and spin-down carriers, $\Delta\mu = (\mu_\uparrow - \mu_\downarrow)$, is called the "spin accumulation" [$\mu_\uparrow$ ($\mu_\downarrow$) = $-eV + E_{F\uparrow}$ ($E_{F\downarrow}$) at 0 K] (see Figure 1d). This polarization diffuses and is measured below electrodes 3 and 4 away from the electrical current (Figure 1b). The spatial spread of the spin polarization in a material is characterized by the spin diffusion length, $l_{sf}$, which is related to the spin–lattice relaxation time, $\tau_{sf}$ (spin lifetime), in that material by $l_{sf} = (D\tau_{sf})^{1/2}$, where $D$ is the diffusion coefficient.

An example of a room-temperature nonlocal spin signal associated with the measurement of the difference in spin accumulation amplitude between these contacts is shown in Figure 1c. The amplitude of the signal depends not only on the length but also on the mean contact resistance, $R_b$, of the tunnel barrier between the graphene and the electrodes. A larger resistance prevents spin escape into the electrodes and preserves a larger spin polarization. The measurement is usually analysed according to a one-dimensional model based on the drift–diffusion equations (see the later section "Analysis of experimental results on graphene" for details). In the notation used in this article, the nonlocal spin signal, $\Delta R_{nl}$, can be expressed as[17]

$$\Delta R_{nl} = \pm \frac{2\gamma^2 \rho_{sq} l_{sf}}{w} \frac{\left(wR_b/\rho_{sq}l_{sf}\right)^2 \exp(-L/l_{sf})}{\left(1+wR_b/\rho_{sq}l_{sf}\right)^2 - \exp(-2L/l_{sf})} \qquad (1)$$

where $\gamma$ is the injection polarization, $\rho_{sq}$ is the square (sheet) resistance of the material (graphene in this case), $l_{sf}$ is its spin diffusion length, $w$ is its width, and $L$ is the distance between injection and detection.

This expression predicts an exponential decay, $\exp(-L/l_{sf})$, of the spin signal as a function of the device length, except when $l_{sf}$ is very long, leading to slower $1/L$ decay. Based on the experimentally observed exponential variation, a spin diffusion length of 1.6 μm was derived for the graphene sample.[17]

The spin-transport parameters can also be extracted using the Hanle effect, in which a magnetic field is applied in a direction perpendicular to the spin accumulation and causes precession and dephasing of the spins in the lateral channel (see Figure 2a). This eventually leads to an oscillating decay of the spin signal, as illustrated by the experimental results of Yang et al.[20] in Figure 2b–c. When the alignment between the applied field and the spin polarization suppresses the precession effects, the spin signal is restored. Fitting of the Hanle curves with solutions of Bloch equations leads to the determination of the diffusion constant $D$, spin lifetime $\tau_{sf}$, and spin diffusion length $l_{sf}$. For their graphene samples, Popinciuc et al. found $\tau_{sf}$ and $l_{sf}$ values of up to 0.2 ns and 2.2 μm, respectively, at room temperature. Values in the same range were also later found by the Kawakami group.[19,21–23] Indeed, using the nonlocal techniques, they found $\tau_{sf} \approx 1$ ns at 4 K and $\tau_{sf} \approx 0.3$ ns at room temperature. Figure 3 shows the local and nonlocal signals at 4 K and room temperature, illustrating the opposite signs of the two types of signal.

*Recent results on large-scale graphene*

Whereas most of the spin-transport measurements discussed so far were made on exfoliated graphene on $SiO_2$, recent publications have also reported experiments performed on graphene grown by chemical vapor deposition (CVD) on copper foils and transferred onto $SiO_2$.[24] These experiments showed spin-transport properties similar to those of exfoliated graphene and introduced the interesting possibility of large-scale production of spin-transport devices, because large areas of CVD graphene can be fabricated easily.

Another very interesting alternative for large-scale integration is epitaxially grown graphene on silicon carbide (SiC). In addition to a large size, epitaxial graphene (EG) samples also show very high mobility[25] (see also the articles in this issue by Ruan et al. and Nyakiti at al.). However, graphene layers grown on the silicon-terminated face (Si face) of SiC show different structures and properties from those grown on the carbon terminated face (C face), and this is also true for spin transport.

For the Si face, which allows easier control of layer growth, Maassen et al.[26] performed nonlocal spin transport measurements on multilayer epitaxial graphene (MEG). They found an average mobility of ~1900 $cm^2V^{-1}s^{-1}$ and $\tau_{sf}$ values on par with those of the best exfoliated samples (up to 2.3 ns at low temperature) but with surprisingly small diffusion constants and $l_{sf}$ values in comparison.

For the C face, Dlubak et al.[27] explored spin transport on MEG (~10 layers) (see Figure 4). Although the number of layers is more difficult to control, this type of graphene is composed of uncoupled monolayer graphene sheets (it is not simply a thin graphite layer), leading to better transport properties. They found very high mobilities of ~17,000 cm$^2$V$^{-1}$s$^{-1}$. In these samples, cobalt/alumina tunnel junctions of very large resistance, in the megaohm range, were used as injectors and detectors.[28]

Local magnetoresistance (MR) curves obtained with MEG[27] are shown in Figure 4c with local spin signals ($\Delta R$) in the megaohm range. These observed spin signals, much larger than the resistances of the cobalt electrodes and graphene channel, are the largest spin signals ever observed with graphene. In Figure 5b, the variation $\Delta R/R \propto 1/LR_b$ identifies the regime expected for a very large tunnel barrier resistance compared to the spin resistance of the graphene channel (defined as $R_{ch}^s = \rho_{sq}l_{sf}/w$) and the electrode resistances and a very large $l_{sf}$ value compared to $L$. The corresponding physics depends on the ratio between the electron dwell time (also called transit time) which is proportional to $LR_b$ and $\tau_{sf}$. The impressive spin signals of these devices can be explained only by $l_{sf}$ in the 100-μm range and above. Such spin diffusion lengths, much longer than those reported previously, are probably related to the very high mobility and high quality of C-face SiC epitaxial graphene.

*Spin relaxation in graphene*

Considering spin relaxation mechanisms, it was observed by the van Wees group[29] that $l_{sf}$ increased linearly with the diffusion constant (proportional to $\tau_p$, the momentum scattering time). This led the authors to suggest that, for their samples at least, the mechanism of spin relaxation was of the Elliot–Yafet (EY) type.[30] This mechanism predicts that, for each carrier momentum scattering event, there is a small probability (related to spin–orbit coupling) of spin flip and hence spin information loss. Thus, $\tau_{sf}$ is expected to increase with $\tau_p$.

Another interesting result was more recently obtained for MLG by the same group.[31] They found that $\tau_{sf}$ increased with the number of layers. In the framework of the EY mechanism,[30] this increase in $\tau_{sf}$ can be attributed to better screening of the external scattering potentials, as reported for suspended graphene.[32]

In agreement with the conclusions of the van Wees group,[29] the Kawakami group found a decrease in $\tau_{sf}$ as $\tau_p$ decreased, suggesting the dominance of EY[30] spin relaxation. However, interestingly, in experiments on bilayer graphene (BLG), they found the opposite behavior: a $\tau_{sf}$ value of up to 6.2 ns at low temperature and close to 1 ns at room temperature.[22] (Concurrently, a similar $\tau_{sf}$ value of 2 ns at room temperature was also found by Yang et al.;[19] see Figure 2b). Compared to the experiments on SLG by the same group, surprisingly, this inverse dependence on $\tau_p$ suggests the dominance of the Dyakonov–Perel (DP)[33] mechanism, which relates spin flips to the accumulation of lattice-induced precession of the spin between scattering events. Thus, increased $\tau_p$ leads to increased $\tau_{sf}$. Yang et al. also found[23] that organic-ligand-bound gold nanoparticles, although introducing faster momentum scattering by localized charges, had no effect on $\tau_{sf}$.

More recently, the interpretation of the observed variation of $\tau_{sf}$ with $\tau_p$ and its relationship to the two mechanisms has been investigated. In most experiments, the variation of $\tau_p$ (which is proportional to µ, the mobility) is controlled through the variation of the charge density $n$ through a gate, for which $\tau_p \propto 1/n$. However, it was shown[34] that, away from the Dirac point, for the EY mechanism, $\tau_{sf} \propto 1/\tau_p$, and the EY mechanism could lead to DP-like behavior. In addition, it was suggested that the DP mechanism could in some cases lead to EY-like behavior.[35]

As discussed in the following section, however, the value of $\tau_{sf}$ observed in experiments corresponds to the fastest relaxation pathway in the devices. This pathway might not necessarily be in the graphene channel, as relaxation can occur through spin escape to the electrodes, where relaxation is much faster. For example, it was shown that at least a threefold increase in $\tau_{sf}$ could be obtained by replacing pinhole contacts with tunnel contacts, which better isolate the channel from the electrodes.[19]

With contradictory results for $\tau_{sf}$ for different types of samples, the mechanism of spin relaxation in graphene is not yet clear, and it appears that no straightforward distinction can be made between the Elliot–Yafet and Dyakonov–Perel mechanisms with the available experimental data. One direction for future work is the study (both theoretical and experimental) of EG samples for which the longest $\tau_{sf}$ (a few hundred nanoseconds) and $l_{sf}$ (a few hundred micrometers) were found.[27]

## Theoretical model and device physics

We now present a theoretical picture of spin transport in a graphene LSV and show how it can be applied to the experimental determination of $\tau_{sf}$ or $l_{sf}$.

### *General discussion*

The analysis of spin transport developed for metal or semiconductor LSVs has to be adapted to describe similar experiments with graphene (or carbon nanotubes). Because the spin relaxation in graphene is considerably slower than that in metals, a strong relaxation-rate mismatch occurs between the graphene channel and the ferromagnetic electrodes. Hence, to prevent the escape and relaxation of spin accumulation into the electrodes (also called backflow[17,36]) and to obtain a large spin signal, it is necessary to isolate the lateral channel from the electrodes by interface resistances (usually tunnel barriers). However, if the interface resistances are too large, the electron dwell time in the lateral channel becomes longer than $\tau_{sf}$, and the spin signal drops. In the next section, we show that large spin signals occur only within a narrow window of interface resistances.

The spin accumulation, and thus the spin signal, can be preserved even more by considerably reducing the volume available for spin relaxation, that is, by working in geometries where the distance $L$ between the current and voltage contacts is much shorter than $l_{sf}$. Spreading and relaxation of spin accumulation outside the active region can be prevented by working with devices with a confined geometry (see Figure 6).

The propagation of spin currents in lateral devices can generally be described within the framework of the drift–diffusion equations first introduced by van Son et al.[37] and Johnson and Silsbee[38] and then extended by Valet and Fert[39] for the interpretation of giant magnetoresistance, in which the current flow is perpendicular to the layers. Solution of the drift–diffusion equations[40] gives the typical exponentially decaying electrochemical potential profiles $\mu_\uparrow$ and $\mu_\downarrow$ shown in Figure 1 and, thus, $\Delta\mu = (\mu_\uparrow - \mu_\downarrow)$. The spin signal is expressed as $\Delta V$ (or $\Delta R = \Delta V/I$), the difference in voltage (or resistance) between the parallel and antiparallel magnetic configurations, which scales with $\Delta\mu/e$ (or $\Delta\mu/eI$). The amplitude of $\Delta\mu$ is controlled by the balance between the injected spin current (proportional to the current $I$) and the relaxation of the spin accumulation in the whole device, including the channel and, importantly, the magnetic electrodes.

Indeed, the typical situation of a much higher relaxation rate in the electrodes (cobalt, iron, etc.) than in the channel (here, graphene) impacts the device physics. We define the spin resistance of the two-dimensional, nonmagnetic channel as $R_{ch}^s = \rho_{sq} l_{sf}/w$. The corresponding spin resistance of the ferromagnetic electrodes is $R_F^s = \rho_F l_{sf}^F/A$, where $\rho_F$ and $l_{sf}^F$ are the resistivity and spin diffusion length in the electrode, respectively, and A is the cross section of the current flow path. We note that the condition $R_F^s \ll R_{ch}^s$ almost always applies and hence is assumed in the following discussion. The mean contact resistance $R_b$ and barrier spin polarization coefficient $\gamma$ characterize the spin contact resistances $R_\pm = 2(1 \pm \gamma)R_b$.

In a simple physical description, the spin accumulation, $\Delta\mu$, and the spin signal, $\Delta R = \Delta V/I \approx \Delta\mu/eI$, are controlled by $\gamma$, the spin asymmetry coefficient of the interface, and by the ratio $R_b/R_{ch}^s$. This ratio fixes the proportion between the two main relaxation paths: (1) spin escape/backflow to the electrodes, where the spin quickly relaxes, leading to an overall relaxation rate proportional to $1/R_b$ (blue zone in both plots of Figure 7), and (2) intrinsic spin relaxation in the lateral channel, which is proportional to $1/R_{ch}^s$ for the open channels in Figure 6a–b or to $(L/l_{sf})R_{ch}^s$ for the confined channels in Figure 6c–d (red zone in both plots of Figure 7).

The nonlocal open (NLO), local open (LO), and local confined (LC) (nonlocal confined [NLC] is similar to LC) curves in Figure 7 show what is expected for the variation of the spin signals of the devices in Figure 6 in the aforementioned limit $R_F^s \ll R_{ch}^s$. The curves of $\Delta R = \Delta V/I$ (Figure 7a) and $\Delta R/R^p = \Delta V/V_{bias}^p$ (Figure 7b, where $R^p$ is the resistance between the source and drain in the parallel magnetic configuration) as functions of $R_b/R_{ch}^s$ are shown for $L = l_{sf}/5$.

To explain this behavior, we first consider the simplest geometry, LC, shown in Figure 6c, consisting of two contacts and confined spin accumulation (within $L$). The expressions for $\Delta R$ and $\Delta R/R^p$ (in the limit $R_F^s \ll R_b$) are[27,40]

$$\Delta R = \frac{4\gamma^2 R_b}{2\cosh\left(\frac{L}{l_{sf}}\right) + \left(\frac{R_b}{R_{ch}^S} + \frac{R_{ch}^S}{R_b}\right)\sinh\left(\frac{L}{l_{sf}}\right)} \qquad (2)$$

and

$$\frac{\Delta R}{R^p} = \frac{\gamma^2}{1-\gamma^2} \frac{2}{2\cosh\left(\frac{L}{l_{sf}}\right) + \left(\frac{R_b}{R_{ch}^S} + \frac{R_{ch}^S}{R_b}\right)\sinh\left(\frac{L}{l_{sf}}\right)} \qquad (3)$$

These equations give a steplike curve for $\Delta R$, as seen in Figure 7a (curve LC), and a bell-like curve for $\Delta R/R^p$, as seen in Figure 7b (curve LC), as functions of $\log(R_b/R_{ch}^S)$. Three zones can be defined: $R_b \ll (L/l_{sf})R_{ch}^S$, $(L/l_{sf})R_{ch}^S < R_b < (l_{sf}/L)R_{ch}^S$, and $R_b > (l_{sf}/L)R_{ch}^S$.

On the left (tinted blue), for $R_b \ll (L/l_{sf})R_{ch}^S$ [which is the conventional channel resistance, i.e., $R_{ch} = (L/l_{sf})R_{ch}^S$], the interface resistance, $R_b$, is too low, corresponding to the reported "impedance-mismatch" regime. The physical interpretation is that, at low $R_b$, the relaxation occurs mainly through spin escape. In this case, the spin signal is very low, $\Delta R \approx 2\gamma^2 R_b^2/R_{ch}$, and decreases with decreasing $R_b$. Because the device resistance is dominated by the channel resistance, $R^p \approx R_{ch}$, one obtains $\Delta R/R^p \approx 2\gamma^2 R_b^2/R_{ch}^2$.

We next focus on the zone corresponding to the beginning of the steepest slope of the steplike curve of $\Delta R$ (Figure 7a) and the maximum of the bell-like curve of $\Delta R/R$ (Figure 7b). This is the regime in which $(L/l_{sf})R_{ch}^S < R_b < (l_{sf}/L)R_{ch}^S$. The physical interpretation of $\Delta R$ at this point is that the increase of the interface resistance $R_b$ progressively reduces the spin escape to the electrodes as $1/R_b$, thus increasing the spin accumulation and $\Delta R$ in proportion to $R_b$ with $\Delta R = 2\gamma^2 R_b$. With both $\Delta R$ and the device resistance, $R^p \approx 2(1-\gamma^2)R_b$ (which holds for $R_b \gg R_{ch}$), increasing in proportion to $R_b$, $\Delta R/R^p \approx (\Delta R/R)_{max} \equiv \gamma^2/(1-\gamma^2)$ becomes practically constant. This corresponds to the peak of $\Delta R/R^p$ in the LC curve in Figure 7b.

The data for carbon nanotubes in Reference 41, represented by crosses in Figure 7b, are approximately in this regime because $\Delta R$ approximately follows the increase in the interface resistance ($R^p$ increasing from 33 MΩ to 150 MΩ), whereas the values of $\Delta R/R^p$ remain in a narrow range between 58% and 72%. It is worth noting that, as explained earlier, in this regime, the spin relaxation mainly occurs through spin escape to the high-relaxation-rate electrodes. Hence, for example, Hanle experiments performed in this range mainly determine $\tau_{sf}$ linked to spin escape to the electrodes and not relaxation in the channel.

We finally focus on the third zone for the curves. This corresponds to the right of the step of the $\Delta R$ curve and the right of the $\Delta R/R$ bell curve (crossover and red tint in Figure 7). For $R_b > (l_{sf}/L)R_{ch}^S$, the competition between the two relaxation mechanisms is inverted. The limiting factor for spin accumulation is no longer spin relaxation by spin escape ($\propto 1/R_b$) but rather spin relaxation in the length $L$ of the channel [$\propto (l_{sf}/L)R_{ch}^S$]. Then, as seen in Figure 7a, $\Delta R$ progressively saturates at the value of $\Delta R = 4\gamma^2 R_b/\sinh(L/l_{sf})$. In the limit $l_{sf} \gg L$ of the figure, this leads to $\Delta R = 4\gamma^2 R_{ch}^s (l_{sf}/L)$. Whereas $\Delta R$ saturates, $R^p \approx 2R_b(1 - \gamma^2)$ still increases in proportion to $R_b$, so $\Delta R/R^p$ decreases as $1/R_b$, as shown in Figure 7b (also see Equation 4 in the next section). In the variation as $1/LR_b$, the proportionality to $1/L$ is related to the confinement of the spin relaxation within $L$. It is only after this saturation of $\Delta R$ that the Hanle effect becomes directly related to the spin relaxation inside the channel.

The experimental data on graphene reported in Reference 27, characterized by a decrease in $\Delta R/R^p$ as $1/R_b$ (see Equation 4), are distributed between the second and third zones, as represented by the triangle symbols in Figure 7.

We have presented the most interesting case of $l_{sf} \gg L$. However, this highly favorable case does not correspond to many earlier experiments, where $l_{sf}$ was found to be on the order of $L$. As $L$ increases and approaches $l_{sf}$, the range $(L/l_{sf})R_{ch}^s < R_b < (l_{sf}/L)R_{ch}^s$ of the ideal regime (top of the bell curve) shrinks. In other words, the window of large MR in Figure 7 progressively narrows. The cosh and sinh functions in Equations 2 and 3 can no longer be approximated as 1 and $L/l_{sf}$, respectively. This restores the less favorable exponential variations of $\Delta R$ as a function of $L/l_{sf}$ that are usually reported[42] in place of the linear dependences on $L$, reflecting the balance between the spin relaxation in the channel $L$ and the spin escape of Equation 4, for example.

We now turn to the open (unconfined) configurations in Figure 6a–b (curves NLO and LO in Figure 7), in which the spin accumulation/relaxation is not confined between the electrodes but spreads a distance $l_{sf}$ on both sides of the device. In this case, the channel relaxation rate is no longer driven by $(L/l_{sf})R_{ch}^s$ (the ratio $L/l_{sf}$ arises from the confinement over approximately $L$) but rather exhibits the higher rate of $1/R_{ch}^s$, as generally $L/l_{sf} \ll 1$ or, at most, $L/l_{sf} \approx 1$. Hence, as one can see for curves LO and NLO in Figure 7, the crossover to the regime controlled by relaxation inside the channel occurs earlier as a function of increasing $R_b$ at about $R_b/R_{ch}^s \approx 1$. Accordingly, the level of the saturation plateau now scales with $R_{ch}^s$ instead of $(L/l_{sf})R_{ch}^s$. We found saturation at $\Delta R = \gamma^2 R_{ch}^s$ for curve NLO (configuration in Figure 6a) and $\Delta R = 2\gamma^2 R_{ch}^s$ for curve LC (configuration in Figure 6c), in agreement with the standard expressions of Takahashi and Maekawa[42] (or the relevant limit of Equation 1). This is well below the saturation level $\Delta R \approx 4\gamma^2 R_{ch}^s l_{sf}/L$ of curve LC (configuration in Figure 6c) for a confined geometry. It is then not surprising to find that the previously reported $\Delta R$ values were much smaller than the $\Delta R$ values expected from a confined geometry (curve LC), in the device-favorable situation of $l_{sf} \gg L$. This is also true for the nonlocal confined (NLC) case in Figure 6d. For example, when the distance between the outer contacts is $3L$, the spin signal $\Delta R$ (not represented in Figure 7) still saturates to $\Delta R \approx \gamma^2 R_{ch}^s l_{sf}/L$, well above the saturation level of the nonconfined LSVs,[37] again showing the amplification factor of $l_{sf}/L$ due to the confinement. An example of similar amplification by confinement can be found in Reference 43 for a metallic LSV.

As we demonstrate next, the regime of the LC curve in Figure 7 can be clearly identified in experimental results on epitaxial graphene. Other results, on structures of types NLO and LO in Figure 6a–b, correspond to the behavior illustrated for LO in Figure 7.

*Analysis of experimental results on graphene*

In this section, we analyze the experimental results obtained by Dlubak et al.[27] on high-mobility MLG grown epitaxially on C-face SiC. A schematic of the LC-type device, with a graphene channel in contact with two cobalt electrodes through alumina the devices analyzed, with channel lengths of 0.8 μm or 2 μm and tunnel resistances $R_b$ varying between 3 MΩ and 75 MΩ, are listed in Table I, together with the corresponding experimental results.

As indicated previously, the variation of the spin signals in Table I (represented in Figure 5b) is characteristic of the crossover from a regime in which the relaxation is dominated by the interface barrier to one in which it is dominated by the channel. Here, $\Delta R$ is expected to progressively saturate, and $\Delta R/R$ is expected to drop to zero as $1/LR_b$ in the limit of large $R_b$, as shown by Equation 4 in the next paragraph. This variation is consistent with the experimental results shown in Figure 5. The data from Table I are also represented by the triangle symbols in Figure 7a–b.

The experimental results of Table I and Figure 5 were analyzed in Reference 27 on the basis of the general expression in Equation 3. To obtain better insight into the parameters involved in the analysis, we rewrite the asymptotic limit of this equation at large $R_b$ as

$$\frac{\Delta R}{R^p} = \frac{2\gamma^2}{1-\gamma^2} \frac{l_{sf} R_{ch}^S}{LR_b} = \frac{2\gamma^2}{1-\gamma^2} l_{sf}^2 \frac{\rho_{sq}}{wLR_b} \qquad (4)$$

where the graphene square resistance is $\rho_{sq} \approx 1$ kΩ (from independent measurements on the same graphene layer) and $w$ is the width of the channel. The two unknown parameters are $\gamma$ and $l_{sf}$, with $\Delta R/R^p$ being an increasing function of both. Dlubak et al.[27] assumed the largest value found in previous experiments on cobalt/alumina junctions, namely, $\gamma = 0.32$, to derive a lower bound for $l_{sf}$. The best fits between Equation 3 and the experimental results for the different samples were found for the spin diffusion lengths listed in Table I, all in the 100-μm range or slightly above. Additionally, relative to the maximum magnetoresistance of MR = $\gamma^2/(1-\gamma^2)$ expected for a symmetric double tunnel junction without any spin relaxation in the intervening conducting material, the MR values of samples with smaller values of $R_b$ correspond to an efficiency close to 80%.

It should be emphasized that the $l_{sf}$ values listed in Table I are only lower bounds. Smaller values of the spin injection parameter γ, such as those reported in many experiments with graphene, or use of standard expressions that are valid for non-confined geometries[42] would lead to greater $l_{sf}$ values. Nevertheless, it is also important to rule out potential spurious effects. First, the resistances of the graphene channel and electrodes are in the kiloohm and ohm ranges, respectively; hence, they are several orders of magnitude smaller than the experimentally observed $\Delta R$ values. Second, some contact effect, such as an anisotropic MR effect of the tunnel resistances, would be expected to increase with the tunnel resistance, in contradiction to the observations.

One might also ask whether spin signals could be observed in much longer devices, say, 50 or 100 μm. As explained in the earlier section "Analysis of experimental results on graphene" (also see Figure 7), large values of $\Delta R/R$ can be observed only in the window $(L/l_{sf})R_{ch}^s < R_b < (l_{sf}/L)R_{ch}^s$. When $L$ increases and tends progressively to $l_{sf}$, this window shrinks considerably. For moderate values of $L$ such as $L = l_{sf}/10$, the maximum MR value can still be obtained but at the cost of precise tuning of the interface resistance, which is not always possible experimentally. For $L$ above this range, the maximum value of MR, $\gamma^2/(1 - \gamma^2)$, is no longer achievable even with very precise tuning of the barrier. This is illustrated in Figure 5a, where one can see that, for devices similar to those described in Reference 27 but with $L \approx l_{sf}/2$, a large MR (although not as large as the maximum value) can be obtained only with highly tunable tunnel resistances that are several orders of magnitude smaller than that of the experiment.

We finally discuss the origin of the much greater $l_{sf}$ value found for MEG in Table I, through comparisons with other experiments. One cannot exclude the possibility that the smaller $l_{sf}$ values in some other experiments could be due to lower-quality tunnel barriers. Another origin of the discrepancy could be incorrect interpretation of Hanle measurements when spin relaxation is mainly due to spin escape (backflow) and not to intrinsic spin relaxation in graphene. However, the main difference arises from the superior properties of MEG grown on C-face SiC compared to graphene obtained by chemical vapor deposition or exfoliation. One of the advantages of this type of graphene comes from the screening of substrate induced scattering in the top layers of the MLG. This is also the probable origin of the very high mobility achieved without treatment or suspension of the graphene. The second point is that graphene multilayers are flatter and have less corrugation than monolayers. This should reduce ripple-induced spin-orbit coupling and its contribution to spin relaxation and might also explain the longer $\tau_{sf}$ values found in samples with increasing numbers of layers.

Actually, the 100-μm range for the samples in Reference 27 should not be the upper limit of $l_{sf}$ values in graphene. Improvements in the quality of epitaxial graphene, substrates with lead to greater $l_{sf}$ values and further increase the potential of graphene for large-scale spintronic devices, such as spin-only logic circuits.

## Conclusions

Graphene has been experimentally identified as a pertinent medium for the transport of spin information over macroscopic distances with limited losses, thus enabling further work on more complex spin architectures (see Figure 8). The next step in developing graphene for spintronics applications is a thrilling challenge and concerns gate manipulation of the spin information transported in the graphene channel. However, a pertinent implementation has yet to be identified. Several theoretical predictions have been formulated, and experimental testing of these predictions has begun. A few interesting proposals are interfaces with ferromagnetic materials or impurities,[50,51] interfaces with large spin-orbit or Rashba interactions,[34,52,53] and also structural tuning of magnetic properties (using nanoribbon edges, nanomeshes, etc.).[54–57] Other interesting types of control of electronic properties are also offered by the Dirac character of the electronic states and their associated pseudospin.[58–60]

Experimentalists will now have to test the implementation of several possible concepts of graphene-based spin logic gates, and this will certainly keep them busy for some time. Nevertheless, the prospects are highly promising for green spin-based low-power memory and logic devices, as well as a global graphene-based electronics platform, fitting a "More than Moore" scenario in the short term and augmenting classic scaling of silicon complementary metal-oxide semiconductor (CMOS), as well as a long-term beyond-CMOS vision.

**Table 1**

| Sample | $L$ (μm) | $R^p$ (MΩ) | $R_b$ (MΩ) | $\Delta R$ (MΩ) | MR (%) | $l_{sf}$ (μm) |
|---|---|---|---|---|---|---|
| A | 2 | 136 | 75.8 | 1.5 | 1.1 | 285 |
| B | 2 | 70 | 39 | 0.4 | 0.7 | 160 |
| C | 2 | 29 | 16.2 | 0.35 | 1.2 | 138 |
| D | 2 | 3.8 | 2.1 | 0.12 | 3.4 | 95 |
| E | 0.8 | 5.8 | 3.2 | 0.55 | 9.4 | 246 |

Spin diffusion lengths $l_{sf}$ in the channel of epitaxial graphene lateral spin valves (LSVs) obtained by fitting Equation 3 to experimental results[27] with only l sf as a free parameter. All other parameters (except γ) extracted from sample measurements. The calculations were performed with γ = 0.32, the maximum value of γ found[27] for the spin polarization of tunneling from cobalt through alumina in similar devices, and provide only a lower bound of $l_{sf}$. $L$, device length; $R^p$, device resistance in the parallel magnetic configuration of the cobalt electrodes; $R_b$, interface resistance as defined in the text; $\Delta R$, spin signal (i.e., the absolute change in device resistance between the parallel and antiparallel magnetic configurations); MR = $\Delta R/R^p$, magnetoresistance.

**Figures captions**

**Figure 1.** (a) A four-terminal LSV device based on a single-layer exfoliated graphene flake showing the graphene shape. (b) Nonlocal measurement geometry of a graphene spin-valve device. The current $I$ is injected between electrodes 3 and 4, and a voltage $V$ is detected between electrodes 1 and 2. (c) A nonlocal spin signal appears as a difference in the nonlinear resistance $R_{nl} = V/I$ in different parallel and antiparallel configurations of the magnetic electrodes as the magnetic field $H$ is swept. These measurements were made at room temperature in a 600-nm-wide device with a 3-μm interelectrode spacing between central electrodes 2 and 3. (d) Schematic representation of the spin-dependent chemical potentials $\mu_\uparrow$ and $\mu_\downarrow$ induced by injecting a spin-polarized current from a ferromagnetic electrode into the graphene lateral channel of the device in (b). The difference $\mu_\uparrow - \mu_\downarrow$ is the spin accumulation giving rise to the measured spin signal. The arrows indicate the magnetization directions of the electrodes. The green (red) dots correspond to the chemical potential being probed with the magnetization pointing downward (upward). Adapted from Reference 17.

**Figure 2.** Hanle effect: (a) Application of an external magnetic field perpendicular to the electrode's magnetization (i.e., perpendicular to the channel plane) forces the precession of the polarization of the spin current. Top: Spin current is conserved in the absence of a perpendicular field. Bottom: The applied field is just strong enough to induce one-half precession during channel transport. (b–c) Room-temperature Hanle curves recorded by Han et al.[19] (on nonlocal open devices as shown in Figure 5a) in both the parallel and antiparallel states of electrode magnetization. Oscillating and decaying device resistances as a function of the applied field are observed. The spin lifetime $\tau_{sf}$ and diffusion constants were extracted from fits of the data. The results in (b) reflect direct ferromagnet/graphene contact, whereas in (c), the contact resistance is increased by the insertion of a tunnel barrier. An increase in spin lifetime from 84 ps to 448 ps was obtained when tunnel barriers were used, showing the importance of spin escape to the electrodes (see description in text).

**Figure 3.** (a) Schematic illustration of the measurement configurations for LSVs, contrasting the nonlocal configuration (top), in which the current $I$ is injected between two neighboring electrodes and the voltage $V = IR_{nl}$ is measured at two other electrodes, and the local configuration (bottom), in which the voltage $V = IR$ is measured between the same electrodes in which the current is injected. (b) Typical signals reported for local measurements at 4 K, with observed $\Delta R$ values limited to the 100-Ω range. Reproduced with permission from Reference[19]. ©2012, Elsevier. (c) Similar results obtained at room temperature. Reproduced with permission from Reference[17]. ©2007, Nature Publishing Group. Note that signals that are clear in the nonlocal configuration become noisier in the local configuration for the same sample.

**Figure 4.** LSV device presented in Reference 27: (a) Plan-view scanning electron microscope image of a two-terminal local LSV. The width of the epitaxial graphene (EG) channel (gray) on SiC (blue) is 10 μm, and the distance between the two $Al_2O_3$/Co electrodes (red) is $L = 2$ μm. (b) Optical image of the entire structure, including contact pads. (c) Large local $\Delta R$ spin signals measured at 4 K. Note that the arrows indicate the sweep directions. Reproduced with permission from Reference[27]. ©2012, Nature Publishing Group.

**Figure 5.** (a) Expected evolution of $\Delta R/R$, normalized to the maximum achievable value $\gamma^2/(1-\gamma^2)$, as a function of barrier-to-channel spin resistance ratio, $R_b/R_{ch}^s$, as derived from

Equation 2 in the LC case. Two scenarios are given: $L = l_{sf}/125$ (blue), which corresponds to the range of data in Reference[27] for which the maximum $\gamma^2/(1 - \gamma^2)$ can be achieved, and $L = l_{sf}/2$ (purple), which corresponds to an extrapolation to much larger devices (for the same other parameters) for which the maximum $\gamma^2/(1 - \gamma^2)$ is no longer achievable. The colored oval represents the experimental data range from Reference 27. An example is given of how the efficiency of a representative device drops (dashed arrow) if its length is changed from $l_{sf} = 125L$ (blue circle) to $l_{sf} = 2L$ (purple circle) with all other parameters remaining constant. The solid horizontal arrow shows how one needs to tune $R_b$ precisely to recover a high signal efficiency. The width of the high-efficiency window (the range of resistance) scales with $l_{sf}/L$. (b) Data from Table 1 of Reference 27, shown together with plots of Equation 4 expressed as a function of the product of $R_b$ and $L$ for $l_{sf}$ values of 4 μm (average literature value), 50 μm, 150 μm, and 250 μm.

**Figure 6.** Schematic representations of the different types of LSVs on a graphene (or other) channel: (a) nonlocal detection and open (nonconfined) channel (NLO), (b) local detection and open channel (LO), (c) local detection and confined channel (LC), and (d) nonlocal detection and confined channel (NLC). In all cases, $L$ is the distance between injection electrode and the detection electrode.

**Figure 7.** Expected variations of (a) the total spin signal, $\Delta R$, and (b) the relative spin signal, MR = $\Delta R/R^p$, as functions of the ratio between the interface resistance, $R_b$, and the channel spin resistance, $R_{ch}^s$, for the LSV configurations in Figure 6a–c. The calculation was performed using Equation 2 with $L = l_{sf}/5$. The symbols indicate—schematically, not quantitatively—the experimental results on carbon nanotubes (crosses) and graphene (triangles) discussed in the text. The curves for the NLC LSV of Figure 5 are similar to the LC curves and are not shown. The blue and red colors indicate the zones where the behavior is dominated by spin relaxation by escape to the electrodes and relaxation within the channel, respectively.

**Figure 8.** Comparison of device performance ($\Delta R$) versus channel properties ($l_{sf}$) for different spin-transport media (metals, semiconductors, carbon allotropes), as reported in the literature in the case of simple local or nonlocal lateral-spin-transport devices. The dark gray arrow indicates the increasing $l_{sf}$ requirement for large signals, from spin transistors to more complex spin circuits. Appelbaum et al. also reported interesting results on silicon, with spin-information propagation over distances in the 100-μm range.[48] However, those devices used multiterminal hot-electron structures that allowed the transmission of only a very small fraction ($<10^{-4}$) of the injected current, giving spin signals not comparable with those reported here.[49]

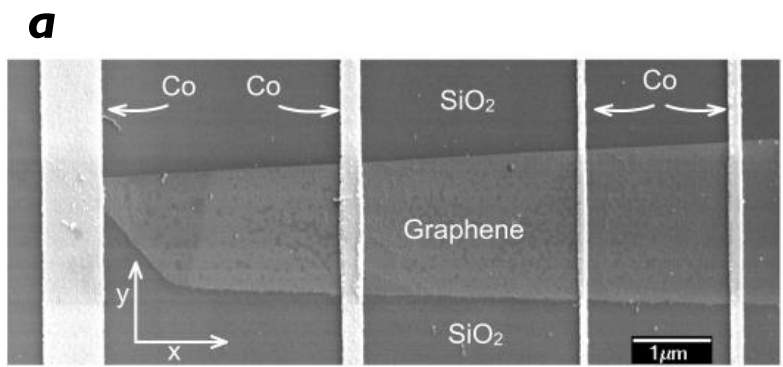
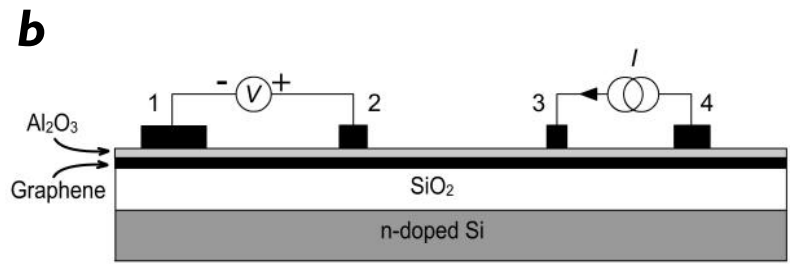
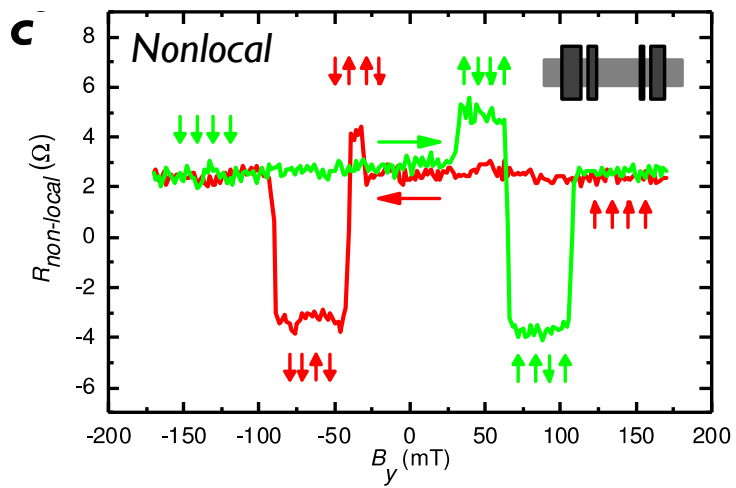
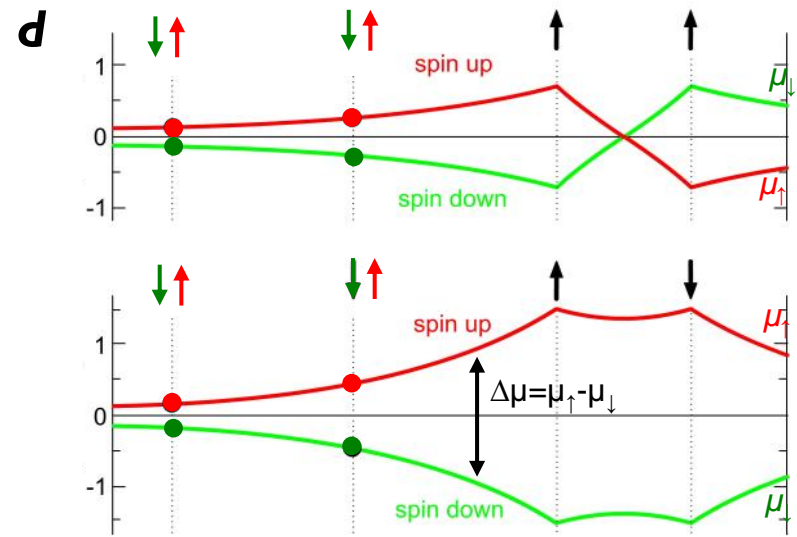

**Figure 1**

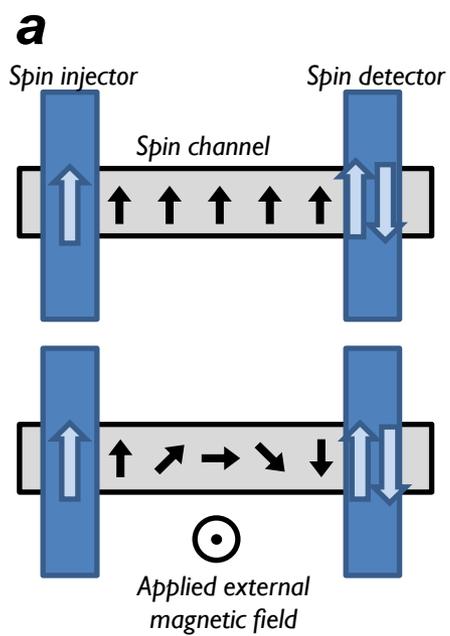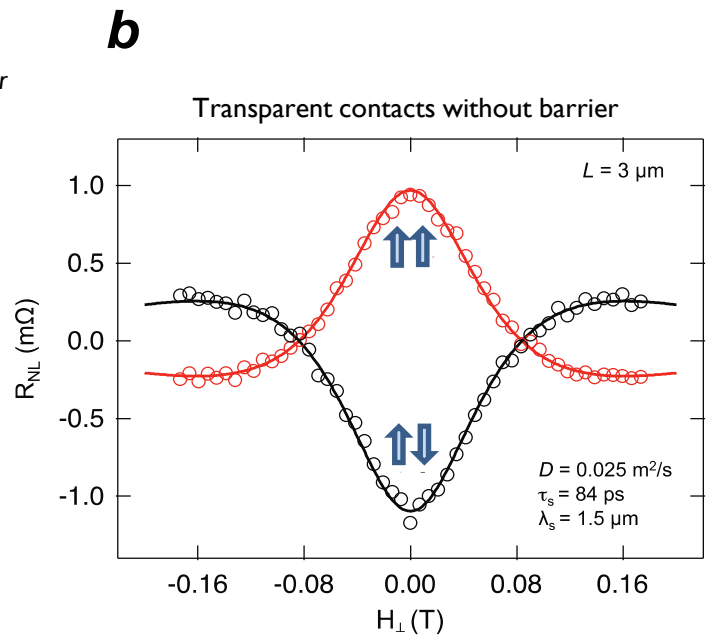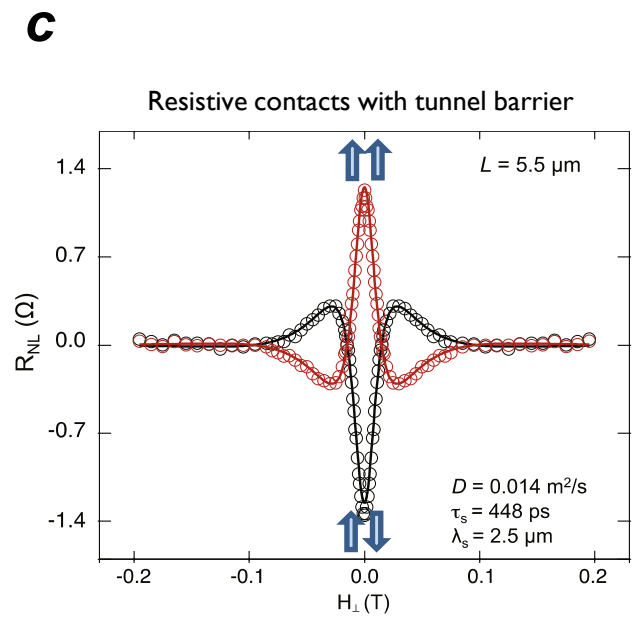

**Figure 2**

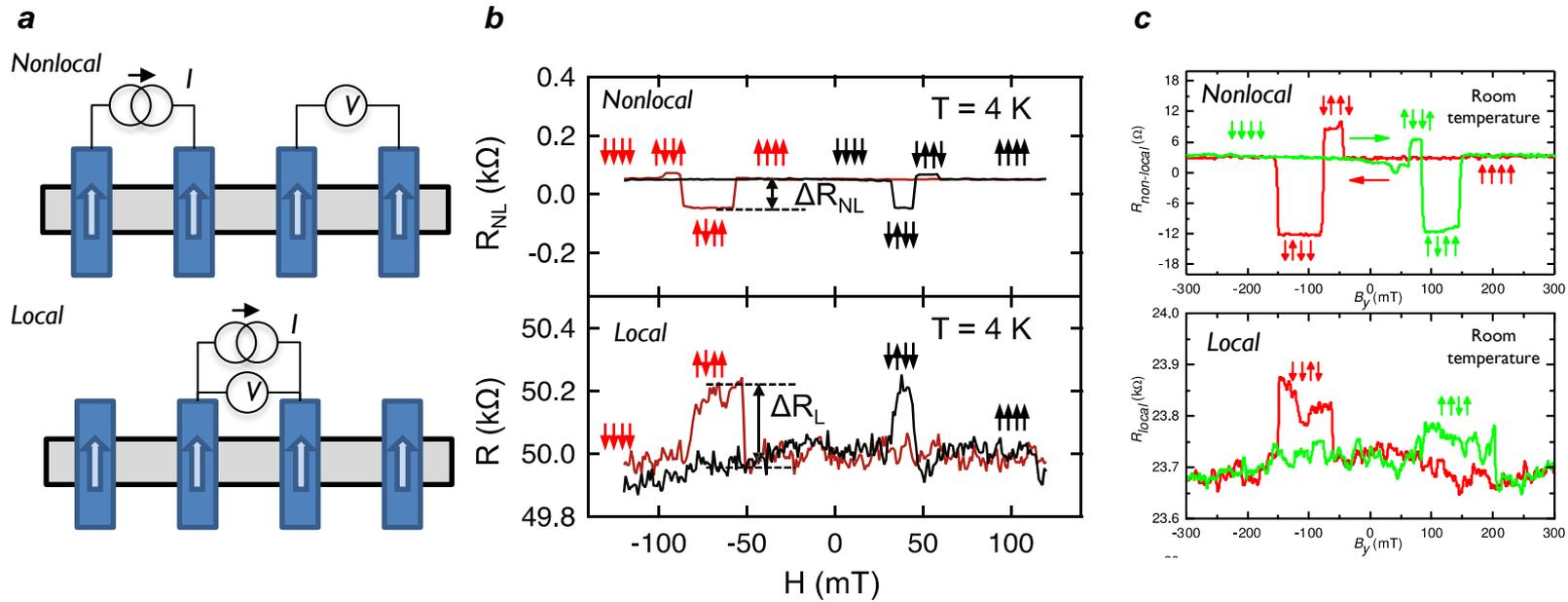

**Figure 3**

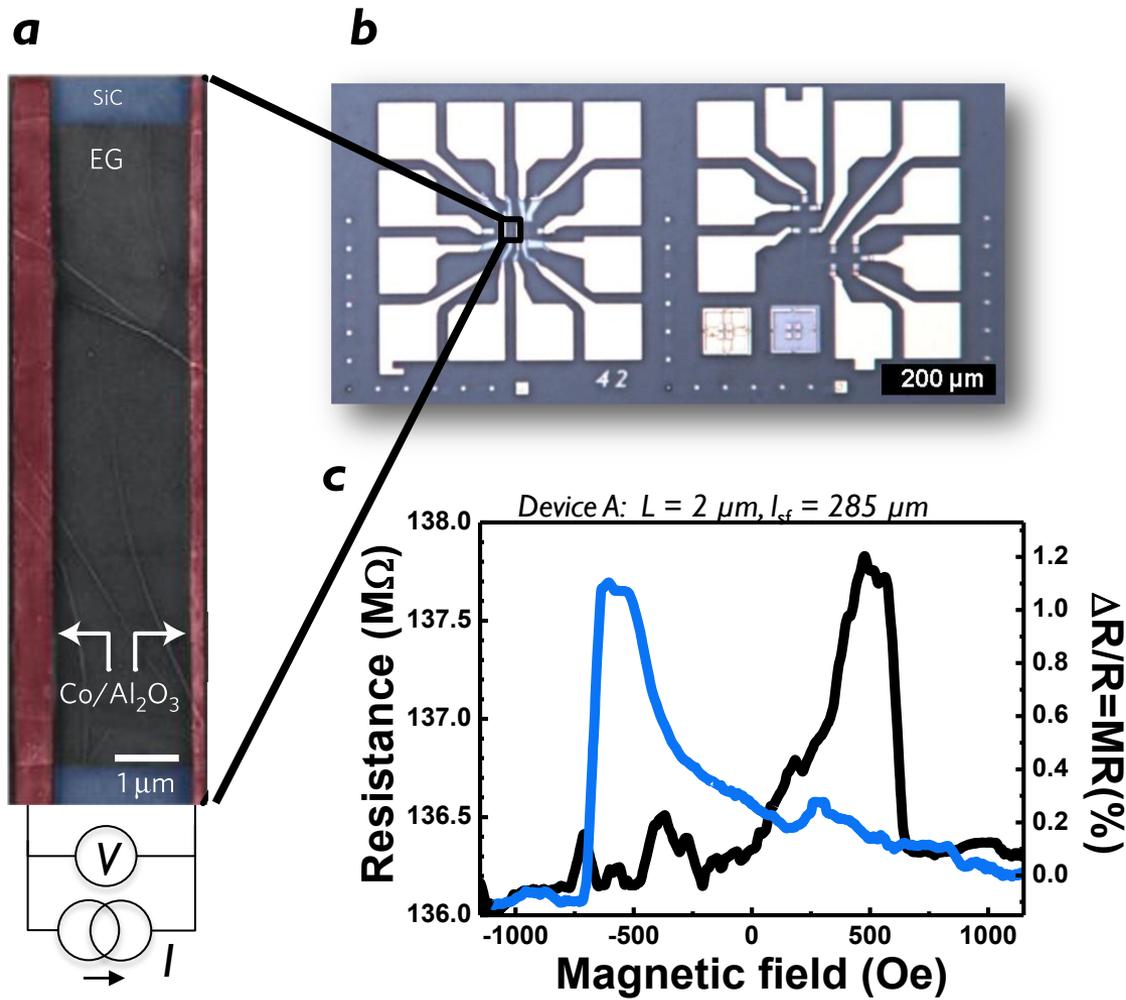

**Figure 4**

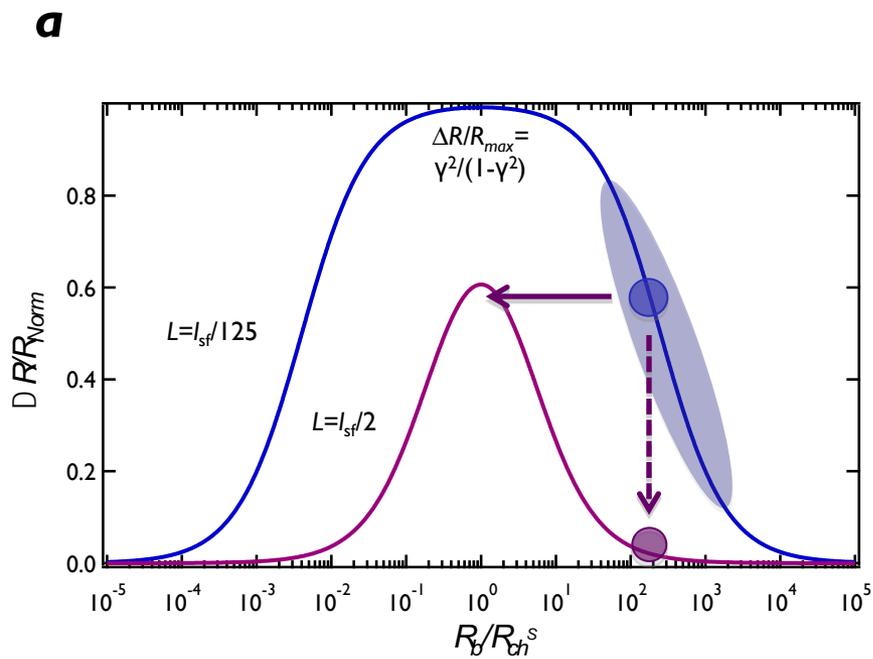 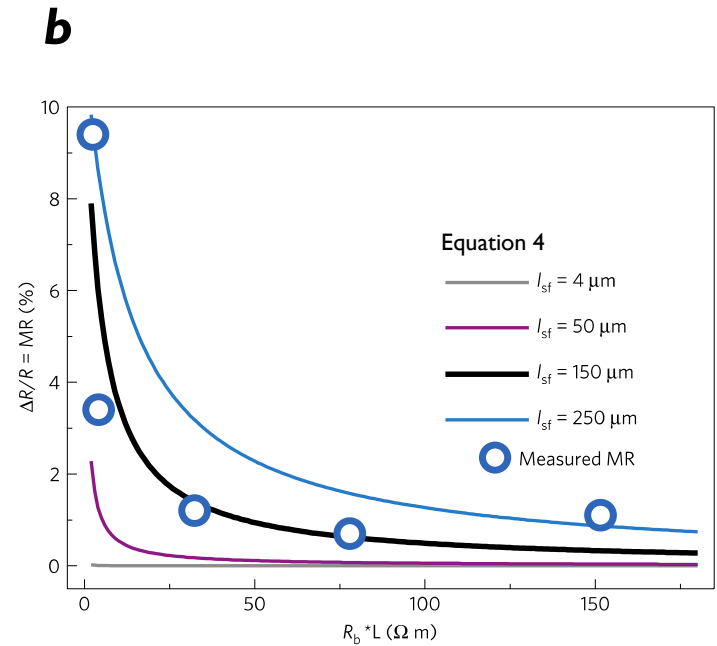

**Figure 5**

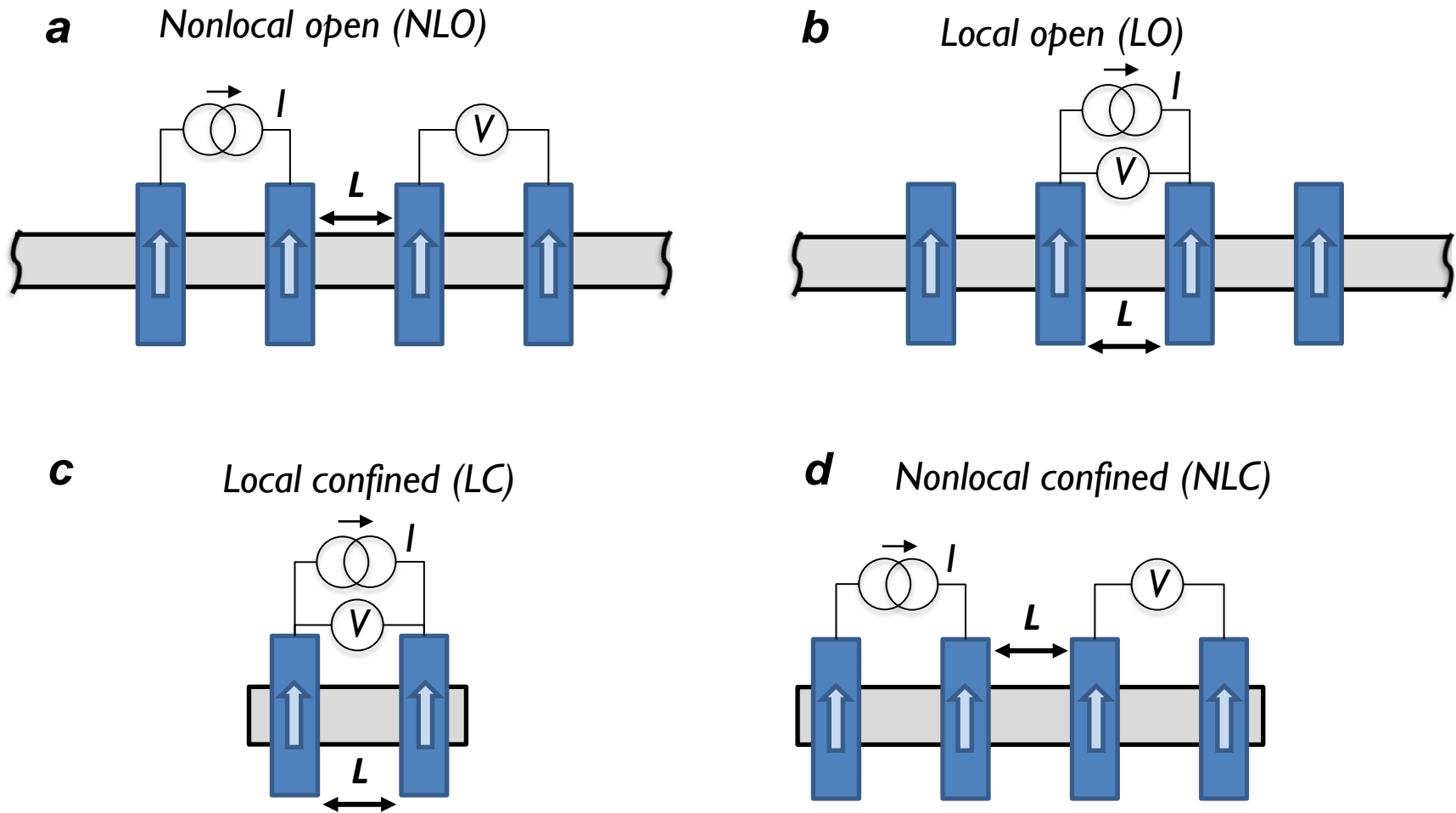

**Figure 6**

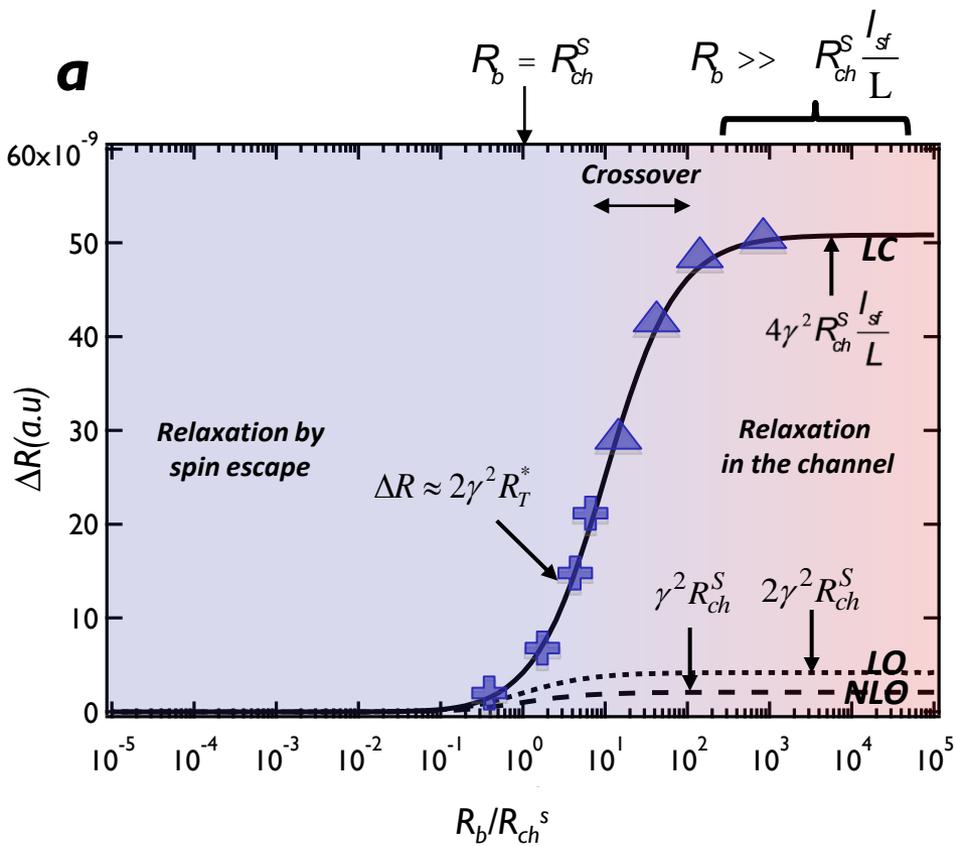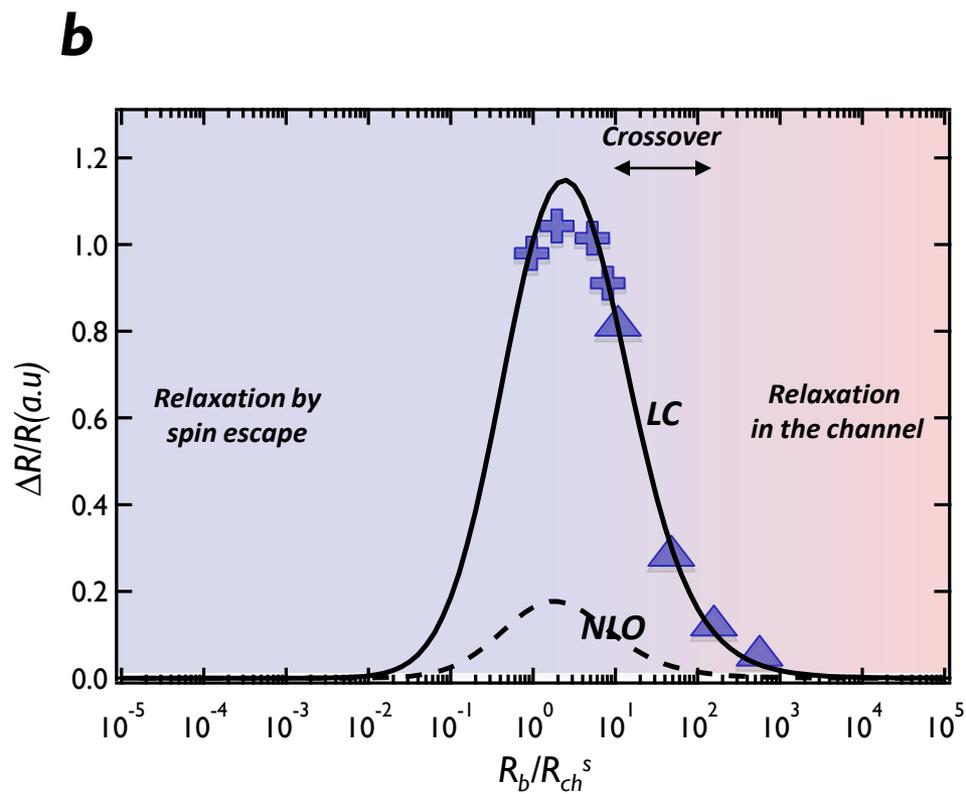

Figure 7

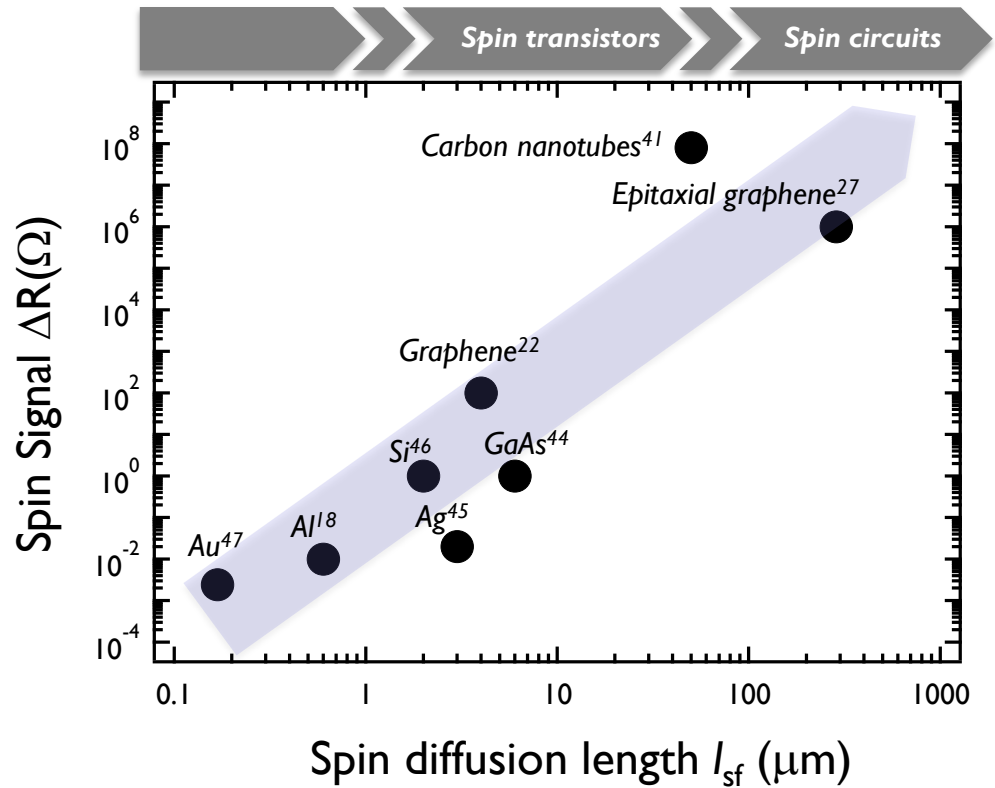

**Figure 8**